\documentclass[prd,aps,floats,preprint,preprintnumbers]{revtex4}
\usepackage{amsmath,amssymb}
\usepackage{graphicx}
\usepackage{graphics}
\usepackage{epsfig}
\usepackage{latexsym,oldgerm}
\def\beq{\begin{equation}}
\def\eeq{\end{equation}}
\def\bea{\begin{eqnarray}}
\def\eea{\end{eqnarray}}
\def\bec{\begin{center}}
\def\eec{\end{center}}
\def\nn{\nonumber}

\begin{document}
\preprint{SU-4252-873}
\preprint{DIAS-STP-08-05}
\title{Constraints from CMB on Spacetime Noncommutativity and Causality Violation}
\author{E. Akofor$^1$}
\email{eakofor@phy.syr.edu}%
\author{A. P. Balachandran$^1$}
\email{bal@phy.syr.edu}%
\author{A. Joseph$^1$}%
\email{ajoseph@phy.syr.edu}%
\author{L. Pekowsky$^1$}%
\email{lppekows@phy.syr.edu}%
\author{B. A. Qureshi$^2$}%
\email{bqureshi@stp.dias.ie}%
\affiliation{\qquad $^1$~Department of Physics, Syracuse University, Syracuse, NY 
13244-1130, USA \\ 
$^2$~School of Theoretical Physics, Dublin Institute for Advanced Studies, 10 Burlington Road, Dublin 4, Ireland\vspace{1cm}}
\begin{abstract}
We try to constrain the noncommutativity length scale of the theoretical model given in \cite{cmbpaper} using the observational data from ACBAR, CBI and five year WMAP. The noncommutativity parameter is not constrained by WMAP data, however ACBAR and CBI data restrict the lower bound of its energy scale to be around $10$ TeV. We also derive an expression for the amount of non-causality coming from spacetime noncommutativity for the fields of primordial scalar perturbations that are space-like separated. The amount of causality violation for these field fluctuations are direction dependent.
\end{abstract}

\maketitle


\section{Introduction} 
In 1992, the Cosmic Background Explorer (COBE) satellite detected anisotropies in the CMB radiation, which led to the conclusion that the early universe was not smooth: there were small density perturbations in the photon-baryon fluid before they decoupled from each other. Quantum corrections to the inflaton field generate perturbations in the metric and these perturbations could have been carried over to the photon-baryon fluid as density perturbations.  We then observe them today in the distribution of large scale structure and anisotropies in the CMB radiation.

Inflation \cite{Starobinsky79, Starobinsky82, Guth81, Linde82, Albrecht82} stretches a region of Planck size into cosmological scales. So, at the end of inflation, physics at the Planck scale can leave its signature on cosmological scales too. Physics at the Planck scale is better described by models of quantum gravity or string theory. There are indications from considerations of either quantum gravity or string theory that spacetime is noncommutative with a length scale of the order of Planck length. CMB radiation, which consists of photons from the last scattering surface of the early universe can carry the signature of spacetime noncommutativity. With these ideas in mind, in this paper, we look for a constraint on the noncommutativity length scale from the WMAP5 \cite{WMAP1, WMAP2, WMAP3}, ACBAR \cite{ACBAR1, ACBAR2, ACBAR3} and CBI \cite{CBI1, CBI2, CBI3, CBI4, CBI5} observational data. 

In a noncommutative spacetime, the commutator of quantum fields at space-like separations does not in general vanish, leading to violation of causality. This type of violation of causality in the context of the fields for the primordial scalar perturbations is also discussed in this paper. It is shown that the expression for the amount of causality violation is direction-dependent. 

In \cite{Sachin}, it was shown that causality violation coming from noncommutative spacetimes leads to violation of Lorentz invariance in certain scattering amplitudes. Measurements of these violations would be another way to put limits on the amount of spacetime noncommutativity.

This paper is a sequel to an earlier work \cite{cmbpaper}. The latter explains the theoretical basis of the formulae used in this paper. In \cite{Queiroz1} another approach of noncommutative inflation is considered based on target space noncommutativity of fields \cite{Queiroz2}.


\section{Likelihood Analysis for Noncommutative CMB} 

The CMBEasy \cite{Doran} program calculates CMB power spectra based on a set of parameters and a cosmological model. It works by calculating the transfer functions $\Delta_{l}$ for multipole $l$ for scalar perturbations at the present conformal time $\eta_{0}$ as \cite{Seljak}
\beq
\label{eq:Delta}
\Delta_l(k, \eta=\eta_0) = \int_{0}^{\eta_{0}} d\eta~S(k,\eta) j_l[k(\eta_0 - \eta)],
\eeq
where $S$ is a known ``source" term and $j_{l}$ is the spherical Bessel function. (Here ``scalar perturbations" mean the scalar part of the primordial metric fluctuations. Primordial metric fluctuations can be decomposed into scalar, vector and second rank tensor fluctuations according to their transformation properties under spatial rotations \cite{brandenberger}. They evolve independently in a linear theory. Scalar perturbations are most important as they couple to matter inhomogeneities. Vector perturbations are not important as they decay away in an expanding background cosmology. Tensor perturbations are less important than scalar ones, they do not couple to matter inhomogeneities at linear order. In the following discussion we denote the amplitudes of scalar and tensor perturbations by $A_s$ and $A_T$ respectively.) The lower limit of the time integral in eq. (\ref{eq:Delta}) is taken as a time well into the radiation dominance epoch. Eq. (\ref{eq:Delta}) shows that for each mode ${\bf k}$, the source term should be integrated over time $\eta$.   

The transfer functions for scalar perturbations are then integrated over ${\bf k}$ to obtain the power spectrum for multipole moment $l$,
\beq
C^{(0)}_{l} = (4\pi^2) \int dk~k^2 P_{\Phi_{0}}(k) |\Delta_l(k,\eta=\eta_0)|^2,
\eeq
where $P_{\Phi_{0}}$ is the initial power spectrum of scalar perturbations (cf. Ref. \cite{cmbpaper}.), taken to be $P_{\Phi_{0}}(k) = A_{s}k^{-3+(n_s-1)}$ with a spectral index $n_{s}$. 

The coordinate functions $\widehat{x}_{\mu}$ on the noncommutative Moyal plane obey the commutation relations
\beq
[\widehat{x}_{\mu}, \widehat{x}_{\nu}] = i \theta_{\mu \nu},~~\theta_{\mu \nu} = -\theta_{\nu \mu} = \textrm{const}.
\eeq
We set $\vec{\theta}^{0} \equiv (\theta^{01}, \theta^{02}, \theta^{03})$ to be in the third direction. In that case, $\vec{\theta}^{0} = \theta~\widehat{\theta}^{0}$ where the unit vector $\widehat{\theta}^{0}$ is $(0, 0, 1)$.

We now write down eq. (79) of \cite{cmbpaper},
\bea
\label{eq:ThetaAngular}
\langle a_{lm} a^{*}_{l'm'}\rangle_{_\theta} &=& \frac{2}{\pi}\int d k \sum_{l''=0, \; l'': \textrm{\tiny{even}}}^{\infty}  i^{l-l'} (-1)^{m}(2l''+1)k^{2} \Delta_{l}(k)\Delta_{l'}(k)P_{\Phi_{0}} (k) i_{l''}(\theta kH)\nn \\
&& \times  \sqrt{(2l+1)(2l'+1)}\left( \begin{array}{ccc}
l & l' & l'' \\
0 & 0 & 0 \end{array} \right)\left( \begin{array}{ccc}
l & l' & l'' \\
-m & m' & 0 \end{array} \right),
\eea
where $i_l$ is the modified spherical Bessel function and $H$ is the Hubble parameter during inflation. In the limit when $\theta = 0$  eq. (\ref{eq:ThetaAngular}) leads to the usual $C_l$'s \cite{Dodelson}:
\beq
C_l=\frac{1}{2l+1} \sum_{m} \langle a_{lm} a^*_{lm}\rangle_0 = (4\pi^2) \int dk~k^2 P_{\Phi_{0}}(k) |\Delta_l(k,\eta=\eta_0)|^2.
\eeq

Our goal is to compare theory with the observational data from WMAP5, ACBAR and CBI. These data sets are only available for the diagonal terms $l=l'$ of eq. (\ref{eq:ThetaAngular}), and for the average over $m$ for each $l$, so we consider only this case. Taking the average over $m$ of eq. (\ref{eq:ThetaAngular}), for $lm = l'm'$ the sum collapses to 
\bea
\label{eq:i-zero}
C^{(\theta)}_l &\equiv& \frac{1}{2l+1} \sum_{m} \langle a_{lm} a^*_{lm}\rangle_\theta =\int dk\, k^2 P_{\Phi_{0}}(k) |\Delta_l(k,\eta=\eta_0)|^2 i_0(\theta k H),\\
C^{(0)}_l &=& C_l . 
\eea

The CMBEasy integrator was modified to include the additional $i_0$ code and the Monte Carlo Markov-chain (MCMC) facility of the program was used to find best-fit values for $\theta H$ along with the other parameters of the standard $\Lambda$CDM cosmology.

In the first run the parameters were fit using a joint likelihood derived from the WMAP5, ACBAR and CBI data.  The outcome of this analysis was inconclusive, as the resulting value was unphysically large.  This result can be understood by examining the WMAP5 data alone and considering a $\chi^2$ goodness-of-fit test, using

\beq
\label{eq:chi-square}
\chi^2 = \sum_l \left(\frac{C^{(\theta)}_{l} - C_{l,data}}{\sigma_l}\right)^2,
\eeq
where $C_{l,data}$ is the power spectrum and $\sigma_l$ is the standard deviation for each $l$ as reported by WMAP observation.

\begin{figure}
\includegraphics[height=11cm]{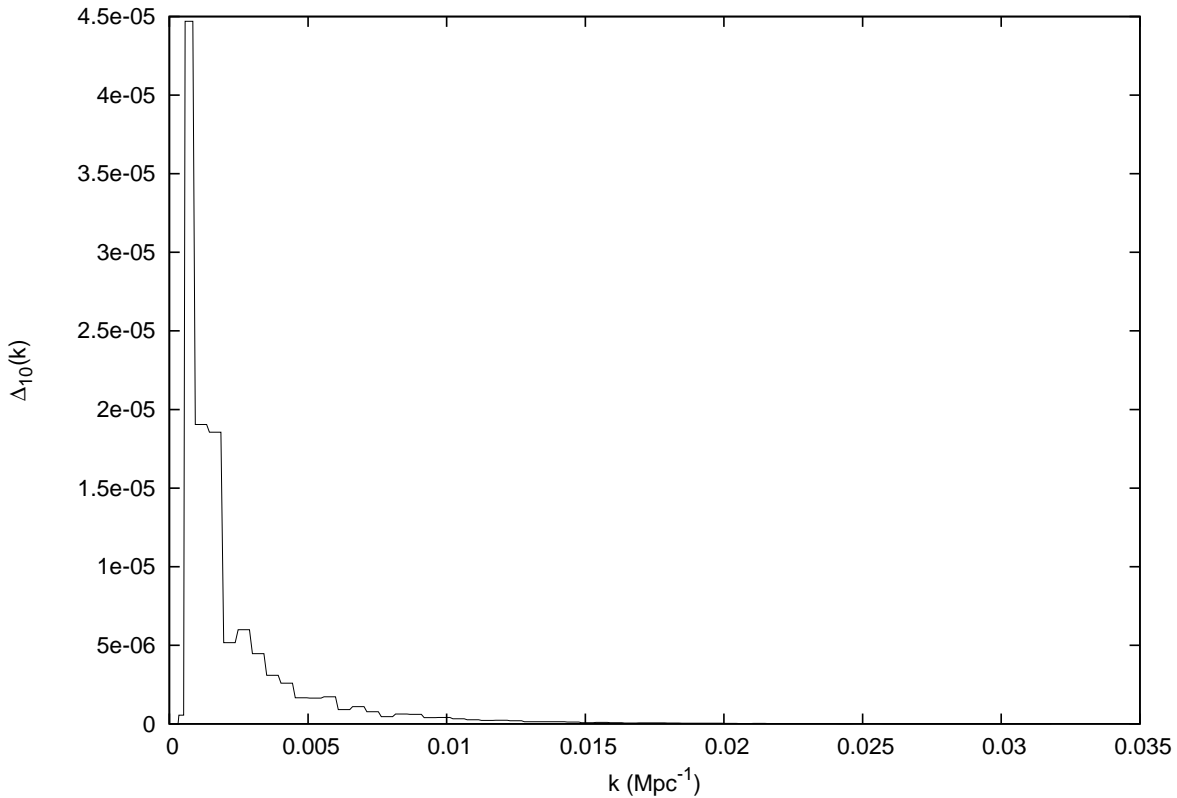}
\caption{Transfer function $\Delta_{l}$ for $l=10$ as a function of $k$. It peaks around $k=0.001$ Mpc$^{-1}$.} \label{fig:1}
\end{figure}

\begin{figure}
\includegraphics[height=11cm]{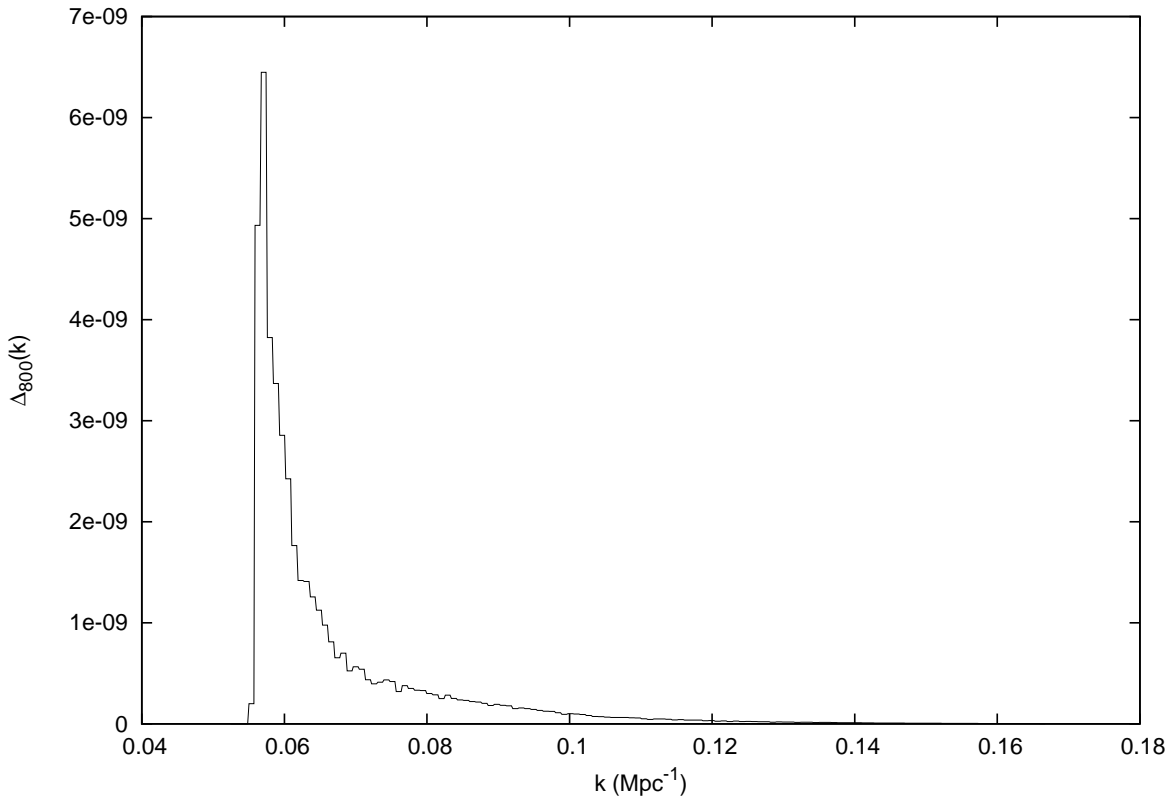}
\caption{Transfer function $\Delta_{l}$ for $l=800$ as a function of $k$. It peaks around $k=0.06$ Mpc$^{-1}$.} \label{fig:2}
\end{figure}

We expect noncommutativity to have a negligible effect on most of the parameters of the standard $\Lambda$CDM cosmology. We therefore consider the effect on the CMB power spectrum of varying only the new parameter $H\theta$. To determine its effect, we consider the shape of the transfer functions $\Delta_l(k)$ as calculated by CMBEasy. The graphs of two such functions are shown in Figs. \ref{fig:1} and \ref{fig:2}. As can be seen, these functions drop off rapidly with $k$, but extend to higher $k$ with increasing $l$. (For example, in Fig. \ref{fig:1}, the transfer function for $l=10$, $\Delta_{10}$, peaks around $k=0.001$ Mpc$^{-1}$ while in Fig. \ref{fig:2}, the transfer function for $l=800$, $\Delta_{800}$, peaks around $k=0.06$ Mpc$^{-1}$.)
As $i_0$ is a monotonically increasing function of $k$ starting at $i_{0}(0)=1$, this means that transfer functions of higher multipoles will feel the effect of noncommutivity first.

\begin{figure}
\includegraphics[height=11cm]{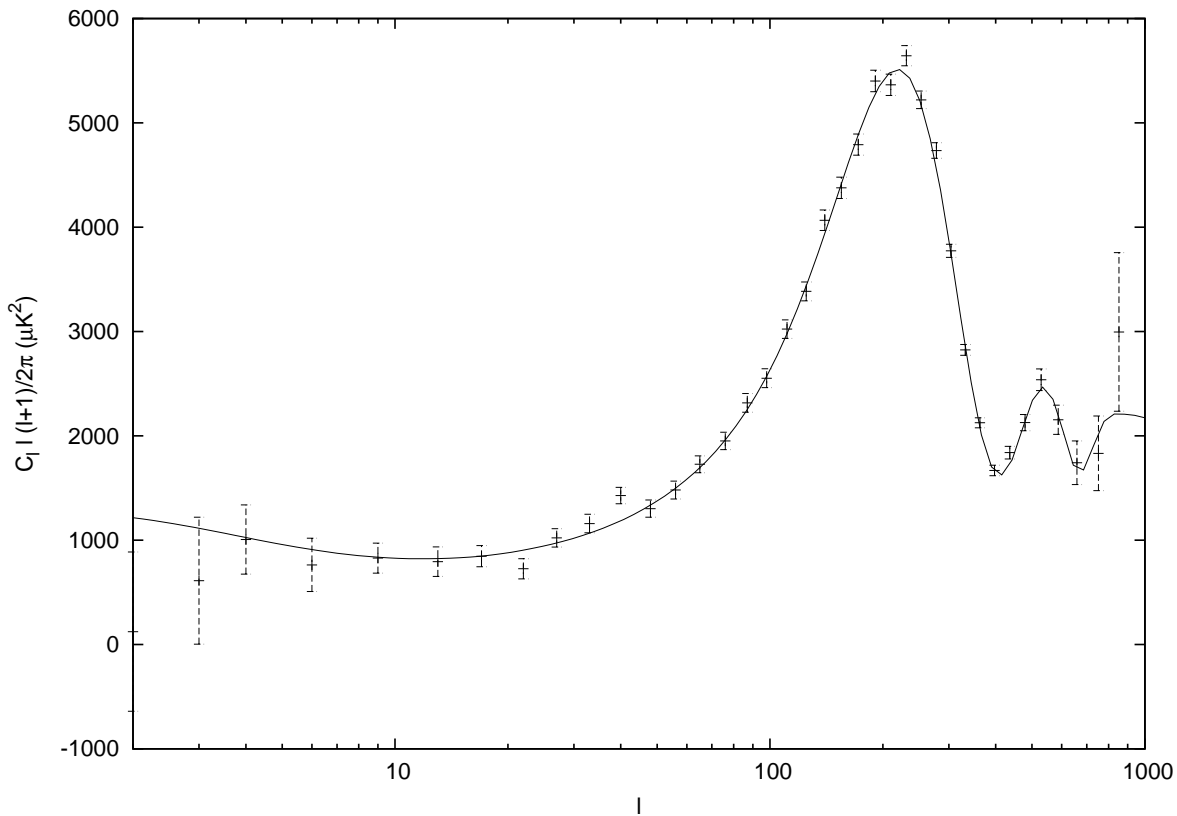}
\caption{CMB power spectrum of $\Lambda$CDM model (solid curve) compared to the WMAP data (points with error bars).} \label{fig:3}
\end{figure}

The spectrum from the WMAP observation is shown in Fig. \ref{fig:3}.  Note in particular that the last data point, corresponding to $l=839$ falls significantly above the theoretical curve. This means that $\chi^2$ can be lowered by a significant amount by using an unphysical value of $H\theta$ to fit this last point, so long as doing so does not also raise adjacent points too far outside their error bars.  Performing the calculation shows that is indeed what happens. We therefore conclude that the WMAP data do not constrain $H\theta$.

\begin{figure}
\includegraphics[height=11cm]{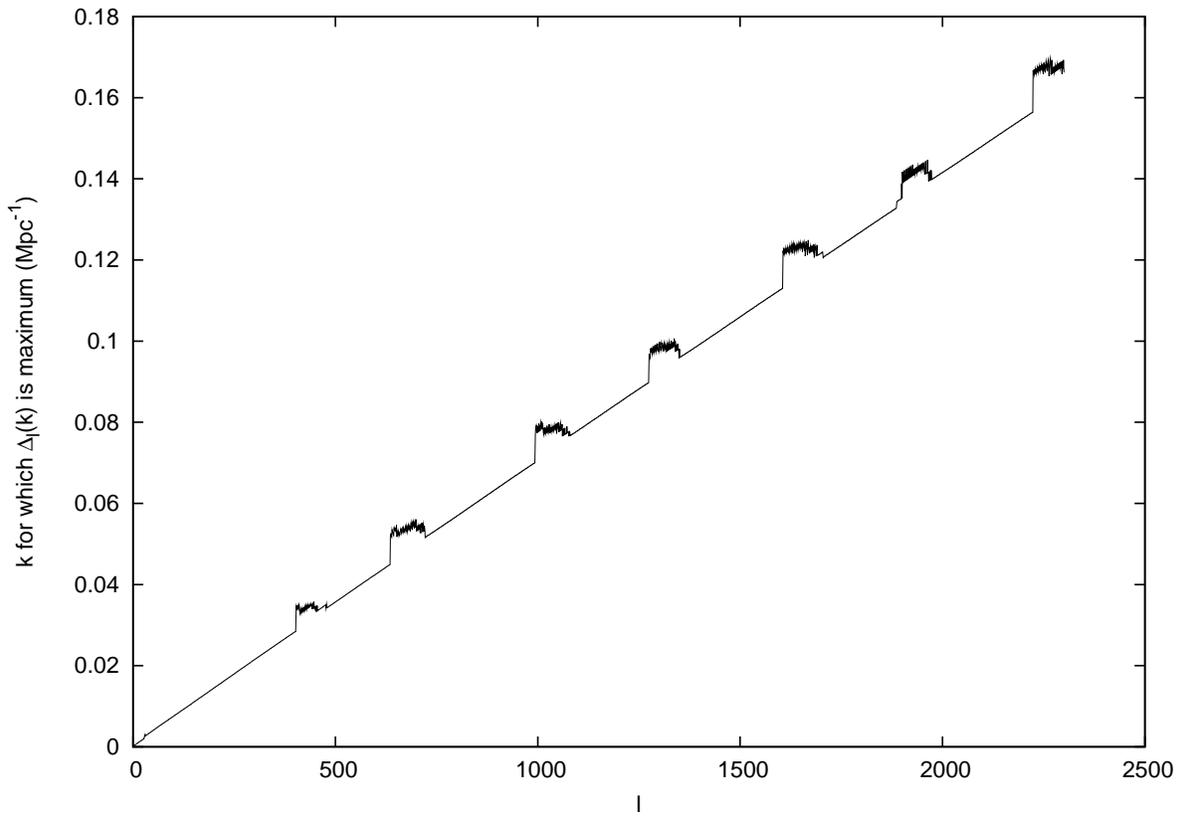}
\caption{The values of $k$ which maximize $\Delta_l(k)$, as a function of $l$} \label{fig:4}
\end{figure}

Fig. \ref{fig:4} shows the values of $k$ which maximize $\Delta_l(k)$, as a function of $l$, which in turn gives a rough estimate of the region over which the transfer functions contribute the most to the integral in eq. (4), and hence the region over which changes in $i_0(H \theta k)$
will most change the corresponding $C_l$. Thus to improve the bound on $H\theta$, we need data at higher $l$ ($l > 839$). In addition, tighter error bars at these higher $l$ will, of course, also help constrain the new parameter.

Based on this analysis we performed a second run of CMBEasy excluding the WMAP data.  This run resulted in a smaller, but still unphysically large, value of $H\theta$.  To see why this happens, we again consider the effect of varying only the new parameter $H\theta$ and examine the behavior of $\chi^2$.

\begin{figure}
\includegraphics[height=11cm]{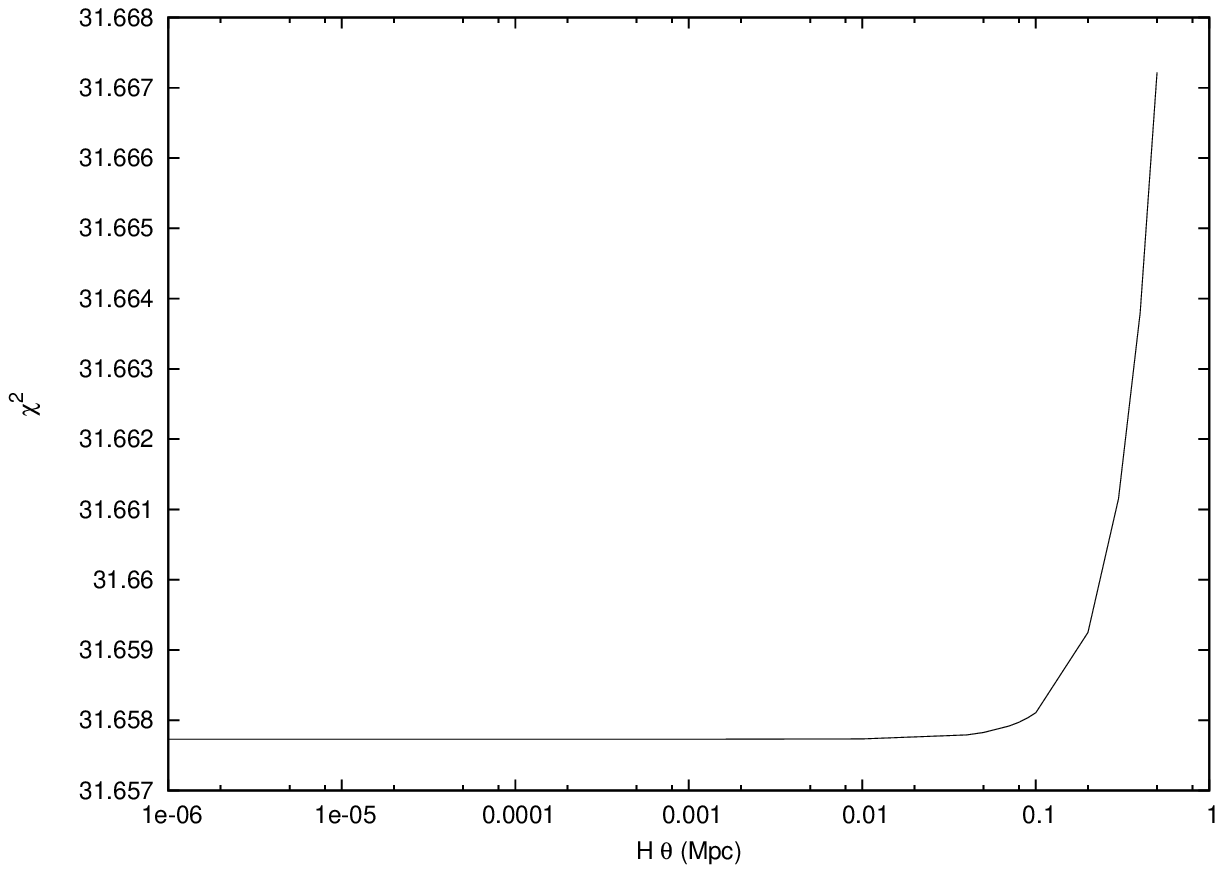}
\caption{$\chi^2$ versus $H\theta$ for ACBAR data} \label{fig:5}
\end{figure}

ACBAR and CBI are CMB data on small-scales (ACBAR and CBI give CMB power spectrum for multipoles up to $l=2985$ and $l=3500$ respectively) and hence may be better suited to determination of $H\theta$.  A plot of $\chi^2$ versus $H\theta$ for ACBAR+CBI data is shown in Fig. \ref{fig:5}.
The plateau between $H \theta = 0$ Mpc and $H \theta = 0.01$ Mpc is not physical, it results from limited numerical precision. Therefore, likelihoods calculated in this range only restrict $H\theta < 0.01$ Mpc and hence cannot indicate whether the best fit is at $H \theta = 0$ Mpc or some small non-zero value.

However, it is possible to put a constrant on the energy scale of spacetime noncommutativity from $H\theta <0.01$ Mpc. We discuss this below.

We can use the ACBAR+WMAP3 constraint on the amplitude of scalar power spectrum $A_{s} \simeq 2.15 \times 10^{-9}$ and the slow-roll parameter $\epsilon < 0.043$ \cite{ACBAR1} to find the Hubble parameter during inflation. The expression for the amplitude of the scalar power spectrum
\beq
A_{s} = \frac{1}{\pi \epsilon}\Big(\frac{H}{M_{p}}\Big)^{2},
\eeq 
where $M_{p}$ is the Planck mass, gives an upper limit on Hubble parameter:
\beq
H < 1.704 \times 10^{-5} M_{p}.
\eeq

On using this upper limit for $H$ in the relation $H \theta <0.01~\textrm{Mpc}$, we have $\theta < 1.84 \times 10^{-9}m^{2}$. 

We are interested to know the noncommutativity parameter at the end of inflation. That is, we should know the value of the cosmological scale factor $a$ when inflation ended. Most of the single field slow-roll inflation models work at an energy scale of $10^{15}$ GeV or larger \cite{Dodelson}. Assuming that the reheating temperature of the universe was close to the GUT energy scale ($10^{16}$ GeV), we have for the scale factor at the end of inflation the value $a\simeq10^{-29}$ \cite{Dodelson}. Thus we have for the noncommutativity parameter, $\sqrt{\theta} < (1.84~a \times 10^{-9})^{1/2} = 1.36 \times 10^{-19}\textrm{m}$. This corresponds to a lower bound for the energy scale of $10$ TeV.


\section{Non-causality from Noncommutative Fluctuations}

In the noncommutative frame work, the expression for the two-point correlation function for the field $\varphi_\theta$ for the scalar metric perturbations contains hermitian and anti-hermitian parts \cite{cmbpaper}. Taking the hermitian part, we obtained the modified power spectrum
\beq
P_{\Phi_{\theta}}({\bf k}) = P_{\Phi_{0}}(k) \; \textrm{cosh}(H\vec{\theta}^{0}\cdot {\bf k}), 
\eeq
where $P_{\Phi_{0}}(k)$ is the power spectrum for the scalar metric perturbations in the commutative case (as discussed in \cite{cmbpaper}), $H$ is the Hubble parameter during inflation. The constant spatial vector $\vec{\theta}^{0}$ is a measure of noncommutativity. The parameter $\theta$ is related to $\vec{\theta}^{0}$ by $\vec{\theta}^{0}=\theta \hat{z}$ if we choose the $z$-axis in the direction of $\vec{\theta}^{0}$, $\hat{z}$ being a unit vector. Also,
\beq
\Phi_{\theta}({\bf k}, t)= \int d^{3}x~\varphi_\theta({\bf x}, t)~\textrm{e}^{-i{\bf k}\cdot {\bf x}}.
\eeq
This modified power spectrum was used to calculate the CMB angular power spectrum for the two-point temperature correlations.

In this section \footnote{This section is based on the work of four of us with Sang Jo. It has been described in \cite{cmbpaper}, but not published.}, we discuss the imaginary part of the two-point correlation function for the field $\varphi_\theta$. In position space, the imaginary part of the two-point correlation function is
obtained from the ``anti-symmetrization" (taking the anti-hermitian part) of the product of fields for a space-like separation: 
\beq 
\label{re:non-causality}
\frac{1}{2}[\varphi_{\theta}({\bf x}, \eta), \varphi_{\theta}({\bf
y}, \eta)]_{-} = \frac{1}{2}\Big(\varphi_{\theta}({\bf x}, \eta)
\varphi_{\theta}({\bf y}, \eta)- \varphi_{\theta}({\bf y}, \eta)
\varphi_{\theta}({\bf x}, \eta)\Big). 
\eeq
The commutator of deformed fields, in general, is nonvanishing for space-like separations. This  type of non-causality is an inherent property of noncommutative field theories constructed on the
Groenewold-Moyal spacetime \cite{Sachin}.

To study this non-causality, we consider two smeared fields localized at ${\bf x}_{1}$ and ${\bf x}_{2}$. (The expression for non-causality diverges for conventional choices for $P_{\Phi_{0}}$ if we do not smear the fields. See after eq. (\ref{non-causal}).) We write down smeared fields at ${\bf x}_{1}$ and ${\bf x}_{2}$.
\bea
&&\varphi(\alpha, {\bf x}_{1}) = \Big(\frac{\alpha}{\pi}\Big)^{3/2}\int d^{3}x~\varphi_{\theta}({\bf x})~e^{-\alpha({\bf x} - {\bf x}_{1})^{2}}, \\
&&\varphi(\alpha, {\bf x}_{2}) = \Big(\frac{\alpha}{\pi}\Big)^{3/2}\int d^{3}x~\varphi_{\theta}({\bf x})~e^{-\alpha({\bf x} - {\bf x}_{2})^{2}},
\eea
where $\alpha$ determines the amount of smearing of the fields. We have
\beq
\lim_{\alpha  \rightarrow \infty}\Big(\frac{\alpha}{\pi}\Big)^{3/2}\int d^{3}x~\varphi_{\theta}({\bf x})~e^{-\alpha({\bf x} - {\bf x}_{1})^{2}}=\varphi_{\theta}({\bf x}_{1}).
\eeq
The scale $1/\sqrt{\alpha}$ can be thought of as  the width of a wave packet which is a measure of
the size of the spacetime region over which an experiment is performed.

We can now write down the uncertainty relation for the fields $\varphi(\alpha, {\bf x}_{1})$ and $\varphi(\alpha, {\bf x}_{2})$ coming from eq. (\ref{re:non-causality}):
\beq
\label{eq:noncausal}
\Delta \varphi(\alpha, {\bf x}_{1}) \Delta \varphi(\alpha, {\bf x}_{2}) \geq \frac{1}{2} \Big| \langle 0 |[\varphi(\alpha, {\bf x}_{1}), \varphi(\alpha, {\bf x}_{2})]|0\rangle \Big|
\eeq

{\it This equation is an expression for the violation of causality due to noncommutativity.}

We can connect the power spectrum for the field $\Phi_0$ at horizon crossing with the commutator of the fields given in eq. (\ref{re:non-causality}):
\beq
\label{eq:commutator}
\frac{1}{2}\langle0|[\Phi_{\theta}({\bf k}, \eta), \Phi_{\theta}({\bf k}', \eta)]_{-}|0\rangle \Big|_{\textrm{horizon crossing}} = (2\pi)^{3}P_{\Phi_{0}}(k)~\textrm{sinh}(H \vec{\theta}^{0}\cdot {\bf k})~\delta^{3}({\bf k}+{\bf k}').
\eeq
Here we followed the same derivation given in \cite{cmbpaper}, using a commutator for the fields to start with, instead of an anticommutator of the fields, to obtain the above result.

The right hand side of eq. (\ref{eq:noncausal}) can be calculated as follows:
\bea
&&\langle 0 |[\varphi(\alpha, {\bf x}_{1}), \varphi(\alpha, {\bf x}_{2})]|0\rangle =\Big(\frac{\alpha}{\pi}\Big)^{3}\int d^{3}x d^{3}y~\langle 0 |[\varphi_{\theta}({\bf x}), \varphi_{\theta}({\bf y})]|0\rangle~e^{-\alpha({\bf x} - {\bf x}_{1})^{2}}e^{-\alpha({\bf y} - {\bf x}_{2})^{2}}\nn \\
&&~~~~~~~~~~=\Big(\frac{\alpha}{\pi}\Big)^{3}\int d^{3}x d^{3}y \frac{d^{3}k}{(2\pi)^{3}} \frac{d^{3}q}{(2\pi)^{3}}~\langle 0 |[\Phi_{\theta}({\bf k}), \Phi_{\theta}({\bf q})]|0\rangle~e^{-i{\bf k}\cdot{\bf x}-i{\bf q}\cdot{\bf y}}e^{-\alpha[({\bf x} - {\bf x}_{1})^{2}+({\bf y} - {\bf x}_{2})^{2}]}\nn \\
&&~~~~~~~~~~=\frac{2}{(2\pi)^{3}}\Big(\frac{\alpha}{\pi}\Big)^{3}\int d^{3}x d^{3}y d^{3}k~P_{\Phi_{0}}(k)\; \textrm{sinh}(H\vec{\theta}^{0}\cdot {\bf k})~e^{-i{\bf k}\cdot({\bf x}-{\bf y})}e^{-\alpha[({\bf x} - {\bf x}_{1})^{2}+({\bf y} - {\bf x}_{2})^{2}]}\nn \\
&&~~~~~~~~~~=\frac{2}{(2\pi)^{3}}\int
d^{3}k~P_{\Phi_{0}}(k)~\textrm{sinh}(H\vec{\theta}^{0}\cdot {\bf
k})~e^{-\frac{{\bf k}^{2}}{2 \alpha} -i{\bf k}\cdot({\bf x}_{1}-{\bf
x}_{2})}. \label{non-causal-commu} 
\eea 
This gives for eq. (\ref{eq:noncausal}), 
\bea 
&&\Delta \varphi(\alpha, {\bf x}_{1}) \Delta
\varphi(\alpha, {\bf x}_{2}) \geq \Big|\frac{1}{(2\pi)^{3}}\int
d^{3}k~P_{\Phi_{0}}(k)~\textrm{sinh}(H\vec{\theta}^{0}\cdot {\bf
k})~e^{-\frac{{\bf k}^{2}}{2 \alpha} -i{\bf k}\cdot({\bf x}_{1}-{\bf
x}_{2})}\Big|.
\label{non-causal} 
\eea 
The right hand side of eq. (\ref{non-causal}) is divergent for  conventional asymptotic
behaviours of $P_{\Phi_{0}}$ (such as $P_{\Phi_{0}}$ vanishing for
large $k$ no faster than some inverse power of $k$) when $\alpha
\rightarrow \infty$ and thus the Gaussian width becomes zero. This
is the reason for introducing smeared fields. 

Notice that the amount of causality violation given in eq. (\ref{non-causal}) is direction-dependent. 

The uncertainty relation given in eq. (\ref{non-causal}) is purely due to spacetime noncommutativity as it vanishes for the case $\theta^{\mu \nu} =0$. It is an expression of causality violation.
\begin{figure}
\includegraphics[height=11cm]{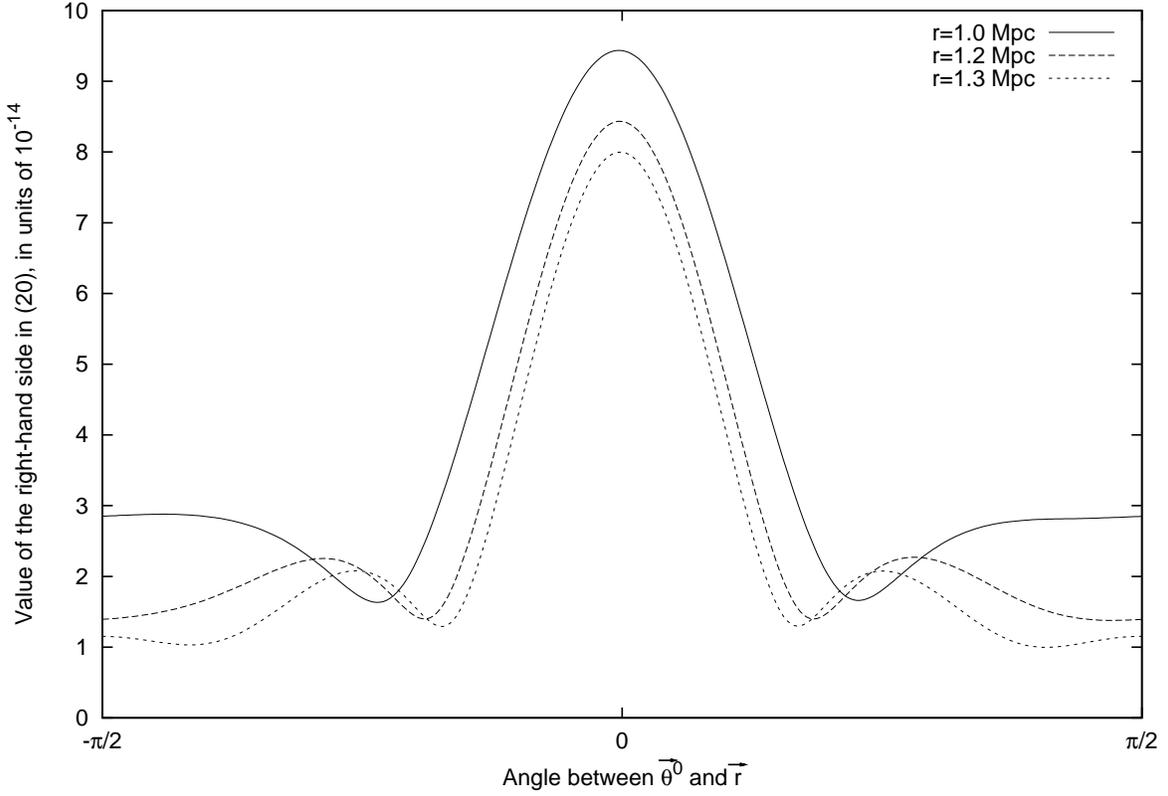}
\caption{The amount of causality violation with respect to the relative orientation between the vectors $\vec{\theta}^{0}$ and ${\bf r}={\bf x}_{1}-{\bf x}_{2}$. It is maximum when the angle between the two vectors is zero. Notice that the minima do not occur when the two vectors are orthogonal to each other. This plot is generated using the Cuba integrator \cite{cuba}.} \label{fig:6}
\end{figure}

This amount of causality violation may be expressed in terms of the CMB temperature fluctuation $\Delta T/T$. We have the relation connecting the temperature fluctuation we observe today and the primordial scalar perturbation $\Phi_{\theta}$,
\bea
\label{eq:deltaToverT}
{\Delta T(\hat{n}, \eta_{0}) \over T} &=& \sum_{l m} a_{l m}(\eta_{0}) Y_{l m}(\hat{n}),\nn \\
a_{lm}(\eta_{0}) &=& 4 \pi (-i)^{l}~\int \frac{d^{3}k}{(2 \pi)^{3}}~\Delta_{l}(k, \eta_{0}) \Phi_{\theta}({\bf k})Y_{lm}^{*}(\hat{k}),
\eea
where $\hat{n}$ is the direction of incoming photons and the transfer functions $\Delta_{l}$ take the primordial field perturbations to the present time $\eta_{0}$. We can rewrite the commutator of the fields in terms of temperature fluctuations $\Delta T/T$ using eq. (\ref{eq:deltaToverT}), but the corresponding correlator differs from the one for the CMB temperature anisotropy. It is not encoded in the two-point temperature correlation functions which as we have seen are given by the correlators of the anti-commutator of the fields.

In Fig. \ref{fig:6}, we show the dependence of the amount of non-causality on the relative orientation of the vectors $\vec{\theta}^{0}$ and ${\bf r}={\bf x}_{1}-{\bf x}_{2}$. The amount of causality violation is maximum when the two vectors are aligned.

\section{Conclusions}
The power spectrum becomes direction dependent in the presence of spacetime noncommutativity, indicating a preferred direction in the universe. We tried a best-fit of the theoretical model in \cite{cmbpaper} with the WMAP data and saw that to improve the bound on $H\theta$, we need data at higher $l$. (The last data point for WMAP is at $l=839$.) We therefore conclude that the WMAP data do not constrain $H\theta$. We also see that tighter error bars at these higher $l$ will also help constrain the noncommutativity parameter. The small-scale CMB data like ACBAR and CBI give the CMB power spectrum for larger multipoles and hence may be better suited for the determination of $H\theta$. ACBAR+CBI data only restrict $H\theta$ to $H\theta < 0.01$ Mpc and do not indicate whether the best fit is at $H \theta = 0$ Mpc or some small non-zero value. However, this restriction corresponds to a lower bound for the energy of $\theta$ of around $10$ TeV. 

Further work is needed before rejecting the initial hypothesis that the other parameters of the $\Lambda$CDM cosmology are unaffected by noncommutivity. It requires performing a full MCMC study of all seven parameters.

Also, we have shown the the existence and direction-dependence of non-causality coming from spacetime noncommutativity for the fields describing the primordial scalar perturbations when they are space-like separated. We see that the amount of causality violation is maximum when the two vectors, $\vec{\theta}^{0}$ and ${\bf r}={\bf x}_{1}-{\bf x}_{2}$, are aligned. Here ${\bf r}$ is the relative spatial coordinate of the fields at spatial locations ${\bf x}_{1}$ and ${\bf x}_{2}$.

\section{Acknowledgements}
We gratefully acknowledge discussions with Sang Jo, Cristian Armendariz-Picon and Eric West. In particular, Sang Jo participated in the work on non-causality in section III. (See the arXiv version of \cite{cmbpaper}.) We also very much thank Duncan Brown for the use of the Syracuse University Gravitation and Relativity (SUGAR) cluster and help with simulations. This work was partially supported by the US Department of Energy under grant number DE-FG02-85ER40231. The work of BQ was supported by IRCSET fellowship.


\begin{thebibliography}{99}
\bibitem{cmbpaper} E. Akofor, A. P. Balachandran, S. G. Jo, A. Joseph and B. A. Qureshi, 
JHEP 05 092  (2008), arXiv:0710.5897 [astro-ph].

\bibitem{Starobinsky79} A. A. Starobinsky, JETP Lett. 30:682-685 (1979), Pisma Zh. Eksp. Teor. Fiz. 30:719-723 (1979).

\bibitem{Starobinsky82} A. A. Starobinsky, Phys. Lett. B117:175-178 (1982).

\bibitem{Guth81} A. H. Guth, Phys. Rev. D, 23, 347-356 (1981).

\bibitem{Linde82} A. D. Linde, Phys. Lett., B108, 389-393 (1982).

\bibitem{Albrecht82} A. Albrecht and P. J. Steinhardt, Phys. Rev. Lett., 48, 1220-1223, (1982).

\bibitem{WMAP1} E. Komatsu, {\it et.al}, ApJS (2008) 
arXiv:0803.0547 [astro-ph].  

\bibitem{WMAP2} M. R. Nolta {\it et al.}, ApJS (2008), arXiv:0803.0593 [astro-ph].

\bibitem{WMAP3} J. Dunkley {\it et al.}, ApJS (2008), arXiv:0803.0586 [astro-ph].

\bibitem{ACBAR1} C. L. Reichardt, {\it et al.}, arXiv:0801.1491 [astro-ph].

\bibitem{ACBAR2} C. L. Kuo {\it et al.}, 
Astrophys. J. 664:687-701 (2007), arXiv:astro-ph/0611198.

\bibitem{ACBAR3} C. L. Kuo {\it et al.}, 
Astrophys. J. 600:32-51 (2004), arXiv:astro-ph/0212289.

\bibitem{CBI1} B.S. Mason {\it et al.}, 
Astrophys. J. 591:540-555 (2007), arXiv:astro-ph/0205384.

\bibitem{CBI2} J. L. Sievers {\it et al.}, 
Astrophys. J. 660:976-987 (2007), arXiv:astro-ph/0509203.

\bibitem{CBI3} J. L. Sievers {\it et al.}, 
Astrophys. J. 591: 599-622 (2003), arXiv:astro-ph/0205387.

\bibitem{CBI4} T. J. Pearson, {\it et al.}, 
Astrophys. J. 591:556-574 (2003), arXiv:astro-ph/0205388.

\bibitem{CBI5} A. C. S. Readhead {\it et al.}, 
Astrophys. J. 609:498-512 (2004), arXiv:astro-ph/0402359.

\bibitem{Sachin} A. P. Balachandran, A. Pinzul, B. A. Qureshi and S. Vaidya, Phys. Rev. D77:025020 (2008), arXiv:0708.1379 [hep-th]; A. P. Balachandran, B. A. Qureshi, A. Pinzul and S. Vaidya, arXiv:hep-th/0608138.

\bibitem{Queiroz1} L. Barosi, F. A. Brito, A. R. Queiroz, 
JCAP 0804:005 (2008), arXiv:0801.0810 [hep-th].

\bibitem{Queiroz2} A. P. Balachandran, A. R. Queiroz, A. M. Marques, P. Teotonio-Sobrinho, 
Phys.Rev.D77:105032 (2008), arXiv:0706.0021 [hep-th].

\bibitem{Doran} M. Doran, 
JCAP 0510:011 (2005), arXiv:astro-ph/0302138.

\bibitem{Seljak} U. Seljak and M. Zaldarriaga, 
Astrophys. J. 469:437-444 (1996), arXiv:astro-ph/9603033.

\bibitem{brandenberger} R. H. Brandenberger, Lect. Notes Phys. 646:127-167 (2004), arXiv:hep-th/0306071.

\bibitem{Dodelson} S. Dodelson, {\it Modern Cosmology}, Academic Press, San Diego (2003).

\bibitem{cuba} T. Hahn, 
Comput. Phys. Commun. 168:78-95 (2005), arXiv:hep-ph/0404043.

\end{thebibliography}
\end{document}